%% file: main.tex
\documentclass[11pt,a4paper]{article}
\usepackage[hyperref]{acl2019}
\usepackage{times}
\usepackage{latexsym}

\usepackage{url}

\usepackage[all]{hypcap}

\usepackage{graphicx}
\usepackage{amsmath}
\usepackage{amssymb}
\usepackage{xspace}
\usepackage{fixfoot}
\usepackage{multirow}

\definecolor{music}{RGB}{176,9,11}
\definecolor{film}{RGB}{117,100,176}
\definecolor{gaming}{RGB}{27,171,72}
\definecolor{tech}{RGB}{0,164,255}
\DeclareMathOperator{\rank}{rank}
\DeclareMathOperator{\jsd}{JSD}
\usepackage{algorithm}
\usepackage{dblfloatfix}
\usepackage{listings}
\usepackage{pgfplots}
\pgfplotsset{compat=1.14}
\usepackage{subfig}
\usepackage{tikz}
\usetikzlibrary{positioning}
\usetikzlibrary{chains}
\usetikzlibrary{calc}
\usepackage{color}

\usepackage{tcolorbox}
\usepackage{todonotes}

\newcommand{\abr}[1]{\textsc{#1}\xspace}

\newcommand{\hidetext}[1]{}

\newcommand\jbg[1]{\textcolor{red}{[ JBG: #1 ]}}

\aclfinalcopy % Uncomment this line for the final submission
\title{Automatic Evaluation of Local Topic Quality}

\author{
    Jeffrey Lund, Piper Armstrong, Wilson Fearn, Stephen Cowley, Courtni Byun, Jordan Boyd-Graber, Kevin Seppi\\
    Computer Science Department\\
    Brigham Young University\\
   {\tt \{jefflund, piper.armstrong, wfearn,}\\\
   {\tt scowley4, emilyhales, kseppi\}@byu.edu}\\
   {\tt jbg@umiacs.umd.edu}
}

\date{}

\begin{document}
\maketitle

\input{acl-2019.tex}

\bibliographystyle{acl_natbib}
\bibliography{main}

\end{document}

%% file: acl-2019.tex
\begin{abstract}
%\jbg{Why consistency? (Should also be in abstract)}
Topic models are typically evaluated with respect to the global topic distributions that they generate, using metrics
such as coherence, but without regard to local (token-level) topic assignments.
Token-level assignments are important for downstream tasks such as classification.
Even recent models, which aim to improve the quality of these token-level topic assignments,
have been evaluated only with respect to global metrics.
We propose a task designed to elicit human judgments of token-level topic assignments.
We use a variety of topic model types and parameters and discover that global metrics
agree poorly with human assignments.

Since human evaluation is expensive we
propose a variety of automated metrics to evaluate topic models at a local level.
Finally, we correlate our proposed metrics with human judgments
from the task on several datasets.
We show that an evaluation based on the percent of topic switches
correlates most strongly
with human judgment of local topic quality.
We suggest that this new metric, which we call consistency,
be adopted alongside global metrics such as topic coherence
when evaluating new topic models.

\end{abstract}

\section{Introduction}
\label{me:sec:intro}

Topic models such as Latent Dirichlet Allocation (or LDA)~\cite{lda} aim to
automatically discover topics in a collection of documents, giving users a
glimpse into themes present in the documents.
LDA jointly derives a set of topics (a distribution over words) and token-topic assignments (a distribution over the topics for each token).
While the topics by themselves are valuable, 
the token-topic assignments are also useful as features for document classification~\cite{labeled-lda,supervised-anchors,labeled-anchors} and, in principle, for topic-based document segmentation.

\iffalse
Topic models such as Latent Dirichlet Allocation (or LDA)~\cite{lda} aim to
automatically discover topics in a collection of documents, giving users a
glimpse into themes present in data.
LDA derives from the data a set of topics,
which are distributions over words.
These topics are then used to assign the specific tokens\jbg{This seems to be about inference, not generative model} in each document
to their corresponding topics, and these assignments are then used to
analyze topical content for each document.
\fi

Given the number of algorithms available for topic modeling, the questions of
algorithm selection and model evaluation can be as daunting as it is important.
When the model is used for a downstream evaluation task
(e.g., document classification),
these questions can often be answered by maximizing downstream task performance.
In most other cases, automated metrics such as topic
coherence~\cite{coherence-automatic} can help assess topic model quality.
%We review the specifics of this and other metrics for topic model evaluation in
%Section~\ref{me:sec:global}.
Generally speaking, these metrics evaluate topic models globally,
meaning that the metrics evaluate characteristics of the topics (word distributions) themselves
without regard to the quality of the topic assignments of individual tokens.

In the context of human interaction,
this means that models produce global topic-word distributions that typically make sense to users
and serve to give a good high-level overview of the general themes
and trends in the data.
However, the local topic assignments can be bewildering.
For example, Figure~\ref{me:fig:split-np} shows typical topic assignments using
LDA.
Arguably, most, if not all, of the sentence should be assigned
to the \underline{Music} topic since the
sentence is about a music video for a particular song.
However, parts of the sentence are assigned to other topics including
\underline{Gaming} and \underline{Technology}, possibly because other
sentences in the same document are concerned with those topics.
Even noun-phrases, such as `Mario Winans' in Figure~\ref{me:fig:split-np},
which presumably should be assigned to the same topic,
are split across topics.

\begin{figure}
    \centering
    \fbox{\begin{minipage}{18em}
    A
    \textcolor{music}{dance\textsuperscript{1} break\textsuperscript{1}} 
    by 
    \textcolor{music}{P.Diddy\textsuperscript{1}}
    is also
    \textcolor{film}{featured\textsuperscript{2}}
    in both
    \textcolor{tech}{settings\textsuperscript{4}}
    of the
    \textcolor{film}{video\textsuperscript{2}},
    \textcolor{music}{intercut\textsuperscript{1}}
    with
    \textcolor{film}{scenes\textsuperscript{2}}
    of
    \textcolor{gaming}{Mario\textsuperscript{3}}
    \textcolor{music}{Winans\textsuperscript{1} playing\textsuperscript{1}}
    the
    \textcolor{music}{drums\textsuperscript{1}}.
    \\
    \\
    \centering
    \begin{tcolorbox}[height=20pt, width=.23\linewidth, bottom=8pt, left=0pt, right=0pt, nobeforeafter, halign=center, valign=center, colframe=black, colback=music]
    {\small Music\textsuperscript{1}}
    \end{tcolorbox}~%
    \begin{tcolorbox}[height=20pt, width=.23\linewidth, bottom=8pt, left=0pt, right=0pt, nobeforeafter, halign=center, valign=center, colframe=black, colback=film]
    {\small Film\textsuperscript{2}}
    \end{tcolorbox}~%
    \begin{tcolorbox}[height=20pt, width=.23\linewidth, bottom=6pt, left=0pt, right=0pt, nobeforeafter, halign=center, valign=center, colframe=black, colback=gaming]
    {\small Gaming\textsuperscript{3}}
    \end{tcolorbox}~%
    \begin{tcolorbox}[height=20pt, width=.27\linewidth, bottom=6pt, left=0pt, right=0pt, nobeforeafter, halign=center, valign=center, colframe=black, colback=tech]
    {\footnotesize Technology\textsuperscript{4}}
    \end{tcolorbox}
    \end{minipage}}
    \caption{Topic assignments from LDA on a sentence from a Wikipedia document. Notice that even
    noun-phrases are split in a way which is bewildering to users. \\}
    \label{me:fig:split-np}
    \fbox{\begin{minipage}{18em}
    A
    \textcolor{music}{dance\textsuperscript{1} break\textsuperscript{1}}
    by 
    \textcolor{music}{P.Diddy\textsuperscript{1}}
    is also
    \textcolor{music}{featured\textsuperscript{1}}
    in both
    \textcolor{film}{settings\textsuperscript{2}}
    of the
    \textcolor{film}{video\textsuperscript{2}},
    \textcolor{film}{intercut\textsuperscript{2}}
    with
    \textcolor{film}{scenes\textsuperscript{2}}
    of
    \textcolor{music}{Mario Winans playing\textsuperscript{2}}
    the
    \textcolor{music}{drums\textsuperscript{2}}.
    \\
    \\
    \centering
    \begin{tcolorbox}[height=20pt, width=.23\linewidth, bottom=8pt, left=0pt, right=0pt, nobeforeafter, halign=center, valign=center, colframe=black, colback=music]
    {\small Music\textsuperscript{1}}
    \end{tcolorbox}~%
    \begin{tcolorbox}[height=20pt, width=.23\linewidth, bottom=8pt, left=0pt, right=0pt, nobeforeafter, halign=center, valign=center, colframe=black, colback=film]
    {\small Film\textsuperscript{2}}
    \end{tcolorbox}~%
    \begin{tcolorbox}[height=20pt, width=.23\linewidth, bottom=6pt, left=0pt, right=0pt, nobeforeafter, halign=center, valign=center, colframe=black, colback=gaming]
    {\small Gaming\textsuperscript{3}}
    \end{tcolorbox}~%
    \begin{tcolorbox}[height=20pt, width=.27\linewidth, bottom=6pt, left=0pt, right=0pt, nobeforeafter, halign=center, valign=center, colframe=black, colback=tech]
    {\footnotesize Technology\textsuperscript{4}}
    \end{tcolorbox}
    \end{minipage}}
    \caption{An example of how topics might be assigned if done by a human.}
    \label{me:fig:natural-tag}
\end{figure}

In the context of downstream tasks, global evaluation ignores the fact that local topic assignments are often used as features.
%For example, topic assignments might be used 
%as features for a topic-based document classification system~\cite{labeled-lda,itm,supervised-anchors}.
If the topic assignments are inaccurate,
the accuracy of the classifier may suffer.

The literature surrounding this issue has focused on improving local
topic assignments, but no metrics that specifically assess the quality of these
assignments have been proposed. Instead the literature evaluates models with global metrics
or subjective examination.

For example, HMM-LDA~\citep{hmm-lda} integrates syntax and topics by
allowing words to be generated from a special syntax-specific topic.
TagLDA~\citep{taglda} adds a tag specific word distribution for each
topic, allowing syntax to impose local topic structure.
The syntactic topic model, or STM~\citep{syntactic-tm}, extends this idea and generates
topics using syntactic information from a parse tree.
An alternative approach to improving local topic quality is by adding
a Markov property to topic assignments.
The hidden topic Markov model~\cite[HTMM]{htmm} does this by adding a switch
variable on each token which determines whether to reuse the previous topic or
generate a new topic.
More recently, \citet{sent-lda} proposed SentenceLDA which assigns each
sentence to a single topic.
CopulaLDA~\citep{cop-lda} supersedes SentenceLDA,
and instead uses copulas to impose topic consistency
within each sentence of a document.

This paper evaluates token-level topic assignment
quality to understand which topic models produce meaningful
local topics for individual documents and proposes metrics
that correlate with human judgment of the quality of these assignments.

%Following the example of previous work on global topic
%evaluation~\cite{coherence-automatic},
%in Section~\ref{me:sec:design}, we propose a user study designed to elicit
%human evaluations of local topic quality.
%In Section~\ref{me:sec:propose} we
%propose a variety of automated metrics designed to perform this evaluation.
%We correlate the task results with our proposed metrics in Section~\ref{me:sec:results}.
%In Section~\ref{me:sec:discuss}, we discuss these results and 
%recommend a new metric of topic model evaluation, which we call
%consistency.

\section{Global Evaluation}
\label{me:sec:global}

Prior work in automated metrics to evaluate topic model quality primarily deals with global evaluations (i.e. evaluations of the topic-word distributions that represent topics).
Early topic models such as LDA were typically evaluated using held-out
likelihood or perplexity~\citep{lda}.
\citet{wallach-eval} give details on how to estimate perplexity.
Indeed, perplexity is still frequently used to evaluate models,
and each of the models mentioned in the previous section,
including CopulaLDA, which was designed to improve local topic quality,
use perplexity to evaluate the model.
However, while held-out perplexity can be useful to test the generalization of
predictive models, it has been shown to be negatively correlated with human
evaluations of global topic quality~\citep{tealeaves}.
This result was elicited using a topic-word intrusion task, in which
human evaluators are shown the top $n$ most probable words in a topic-word
distribution and asked to identify a randomly chosen `intruder' word
which was injected into the word list.
The topic-word intrusion task operates under the assumption that if a 
topic is semantically coherent, then the intruder will be easy to identify.

\subsection{Coherence}

While human evaluation of topic coherence is useful,
automated evaluations are easier to deploy.
Consequently,
\citet{coherence-automatic} propose a variety of automated 
evaluations of topic coherence
and correlate these metrics with human evaluations
using the topic-word intrusion task mentioned above.
They show that an evaluation based on aggregating
pointwise mutual information (PMI) scores across the top $n$ most
likely terms in a topic distribution
correlates well with human evaluations.
This metric, colloquially referred to simply as `coherence',
is currently the most popular form of automated topic model evaluation.
Note that coherence is a measure of global topic quality,
since it considers only the global topic-word distributions.
We follow this pattern of leveraging human intuition in the
development of our own automated metrics proposed in Section~\ref{me:sec:propose}.

%Topic coherence has been well studied.
%For example, through certain types of regularization we can improve topic
%coherence~\cite{coherence-regularize}.
%\citet{coherence-automatic} gave a methodology for automatically performing
%topic-word intrusion tasks.
%Since topic coherence depends on the choice of how many words from each topic
%to consider, work has also been done exploring topic cardinality with respect to
%coherence~\cite{coherence-n}.

\subsection{Significance}
%\jbg{This subsection could be stronger with more on how the long tail affects classification, what sig. misses (motivate your stuff)}
For the purpose of user interaction,
topics are typically summarized by their top $n$ most probable words.
However, when topics are used as features for downstream tasks such as document
classification, the characteristics of the entire distribution
become more important. With this in mind, consider two topics 
which rank the words of the vocabulary by probability in the same order.
Suppose that one of these distributions is more uniform than the other
(i.e., has higher entropy).
While both distributions would be equally interpretable to a human
examining them, the topic-word distribution with lower entropy 
places more weight on the high-rank words and is much more specific.

Using this intuition,
\citet{junk-topic} develops metrics for evaluating topic significance.
While this work was originally used to rank topics, %by significance,
it has also been used to characterize entire models by measuring average
significance across all topics in a single model~\cite{tandem-anchors}.

Topic significance is evaluated by measuring the distance between topic
distributions and some background distribution.
For example,
we can measure significance with respect to the uniform distribution (\abr{SigUni}).
Alternatively,
we can use the empirical distribution of words in the corpus,
which we call the vacuous distribution,
as our background distribution (\abr{SigVac}).

Like coherence, topic significance is a global measure of topic quality since
it considers the topic-word distributions without regard to local topic
assignments.
However, it differs from topic coherence in that it considers the entire
topic distribution.
\citet{tandem-anchors} found that when topics were used as
features for document classification,
models with similar coherence scores
could perform differently on downstream classification accuracy, but
the models with higher significance scores obtained better accuracy.

Automated global metrics have proven useful for evaluating the topics themselves,
that is, the topic-word distributions. 
However, no metric has been shown to effectively evaluate local topic quality.
Therefore, we first correlate existing metrics with human judgment of local topic quality;
we obtain these judgments through the crowdsourcing task described below.

\section{Crowdsourcing Task}
\label{me:sec:design}

Following the general design philosophy in developing the coherence metric in~\citet{coherence-automatic},
we train a variety of models on various datasets to obtain data
with varying token-level topic quality.
We then evaluate these models using crowdsourcing data on a task designed to elicit
human evaluation of local topic model quality.
By then correlating the human evaluation with existing, global metrics,
we determine that global metrics are inadequate, and then propose new metrics
to better measure local topic quality.
%In this section, we first discuss the models we employ,
%then the crowdsourcing task.
%We also discuss the annotator agreement
%with each topic modeling algorithm.

\subsection{Datasets and Models}
\label{me:ssec:models}

We choose three datasets from domains with different writing
styles.
These datasets include Amazon product reviews,%
\footnote{http://jmcauley.ucsd.edu/data/amazon/}
the well known
Twenty Newsgroups dataset,%
\footnote{http://www.ai.mit.edu/people/jrennie/20Newsgroups/}
and a collection of news articles from the New York Times.%
\footnote{http://www.ldc.upenn.edu/Catalog/CatalogEntry.jsp?
          catalogId=LDC2008T19}
We apply stopword removal using a standard list of stopwords,
and we remove any token which does not appear in at least 100
documents.
Statistics for these three datasets can be found in Table~\ref{me:tab:data}.

\begin{table}[ht]
    \small 
    \centering
    \begin{tabular}{l l l l}
    \hline\hline
    Dataset        & Documents & Tokens        & Vocabulary\\
    \hline
    Amazon         & 39388     & 1389171       & 3406  \\
    Newsgroups     & 18748     & 1045793       & 2578  \\
    New York Times & 9997      & 2190595       & 3328  \\
    \end{tabular}
    \caption{Statistics on datasets used in user study and metric evaluation.}
    \label{me:tab:data}
\end{table}

Once again aiming for a wide variety of topic models for our evaluation,
for each of these datasets, we train three types of topic models.
As a baseline, we train Latent Dirichlet Allocation~\cite{lda} on each of the three
datasets using the gensim defaults.%
\footnote{\url{https://radimrehurek.com/gensim}}
CopulaLDA~\cite{cop-lda} is the most recent and reportedly the best model
with respect to local topic quality; we use the authors' implementation and parameters.
Finally, we use the Anchor Words algorithm~\cite{anchors-practical},
which is a fast and scalable alternative to traditional probabilistic topic
models based on non-negative matrix factorization.
In our implementation of Anchor Words we only consider words as candidate anchors if they appear
in at least 500 documents, the dimensionality of the reduced space is 1000, 
and the threshold for exponentiated gradient descent is 1e-10.
By itself, Anchor Words only recovers the topic-word distributions,
so we follow~\citet{supervised-anchors} and use variational inference for LDA
with fixed topics to assign each token to a topic.

\begin{figure*}
    \centering
    \includegraphics[scale=.27]{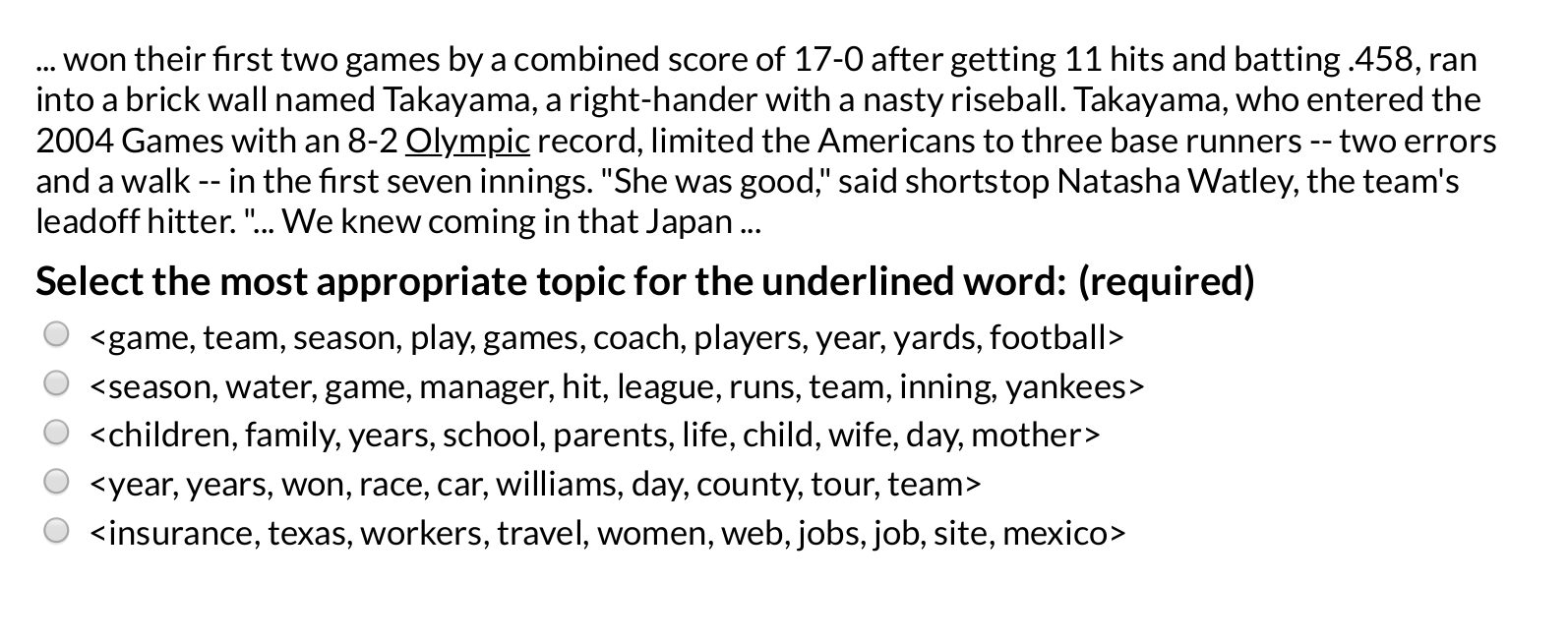}
    \caption{Example of the topic-word matching task. Users are asked to select the topic
    which best explains the underlined token (``Olympic'').}
    \label{me:fig:task}
\end{figure*}

In addition to varying the datasets and topic modeling algorithms, we also vary the number of topics
with the hope of increasing the diversity of observed topic model quality.
For both LDA and Anchor Words, we use 20, 50, 100, 150, and 200 topics.
For CopulaLDA, we use 20, 50, and 100 topics.%
\footnote{Unfortunately, CopulaLDA does not scale beyond 100 topics.
In contrast to LDA and Anchor Words, which run
in minutes and seconds respectively,
CopulaLDA takes days to run using the original authors' implementation.
Our runs with 150 and 200 topics never finished,
as they where finally killed
due to excessive memory consumption on 32GB systems.
}
We vary the number of topics to produce models with small numbers of coherent,
albeit less significant, topics as well as models with large numbers of more
significant topics.
Since each model includes some amount of non-determinism,
we train five instances of each dataset, model, and topic cardinality
and average our results.

In the interest of reproducibility, the data, the scripts for importing and
preprocessing the data, and the code for training and evaluating these topic
models are available in an open source repository.%
% \footnote{https://github.com/jefflund/ankura}.
\footnote{Available after blind review}

\subsection{Task Design}
\label{me:ssec:task}

% Note: We do \textit{not} need an IRB since we are asking
%questions about models. We obtain no information about a living individual. We
%have a quasi-contractual relationship with the crowd source workers to obtain
%the information about those models.

The goal for our crowdsourcing task is to have human annotators
evaluate local topic quality.
Not only will this task allow us to evaluate and compare topic models themselves,
but it will also allow us to determine the effectiveness of automated metrics.
Because local topic quality is subjective,
directly asking annotators to judge assignment 
quality can result in poor inter-annotator agreement.
Instead, we prefer to ask users to perform a task which illuminates the 
underlying quality indirectly.
This parallels the reliance on the word intrusion task to rate 
topic coherence~\citep{tealeaves}.

We call this proposed task `topic-word matching'.
In this task, we show the annotator a short snippet from the data with a single token underlined
along with five topic summaries (i.e., the 10 most probable words in the 
topic-word distribution).
We then ask the user to select the topic which best 
fits the underlined token.
One of the five options is the topic that the model actually
assigns to the underlined token.
The intuition 
is that the annotator will agree more often with
a topic model which makes accurate local topic assignments.
As alternatives to the model-selected topic for the token, we also include the three most
probable topics in the document, excluding the topic assigned to the underlined token.
A model which gives high quality token-level topic assignments
should consistently choose the best possible topic for each individual token, even if these topics are closely related.
Finally, we include a randomly selected intruder topic as a fifth option.
This fifth option is included to help distinguish between an instance where the user
sees equally reasonable topics for the underlined token (in which case, the intruding topic will not be selected), and when there are no reasonable
options for the underlined token (in which case, all five topics are equally likely to be chosen).
Figure~\ref{me:fig:task} shows an example of this task shown to annotators.

For each of our 39 trained models (i.e., for each model type, dataset, and topic
cardinality), we randomly select 1,000 tokens to annotate.
For each of the 39,000 selected tokens, we obtain 5 judgments.
We aggregate the 5 judgments by selecting
the contributor response with the
highest confidence, with agreement weighted by contributor trust.
Contributor trust is based on accuracy on test questions.

We deploy this task on a popular crowdsourcing website%
\footnote{https://www.figure-eight.com}
and pay contributors \$0.12~USD per page, with 10 annotations per page.
For quality control on this task, each page contains one test question.
The test questions in our initial pilot study are questions we
hand-select for their obvious nature.
For our test questions in the final study,
we use the ones mentioned above in addition to questions from the pilot studies
with both high annotator confidence and perfect agreement.
We require that contributors maintain at least a 70\% accuracy on test
questions throughout the job, and that they spend at least 30 seconds per page,
but otherwise impose no other constraints on contributors.
We discuss the results from this final study in 
Section~\ref{me:sec:results}.

\subsection{Agreement Results}
\label{me:ssec:evals}

%The main purpose of our user study is to explore how automated evaluation techniques 
%correlate with human judgments of local topic quality.
%However, the results of our user study are interesting in their own right,
%so we briefly discuss them here.

We first measure inter-annotator agreement using Krippendorff's alpha
with a nominal level of measurement~\cite{kripps-alpha}.
Generally speaking, $\alpha=1$ indicates perfect reliability,
while $\alpha<0$ indicates systematic disagreement.
Over all the judgments we obtain, we compute a value of $\alpha=0.44$, which indicates a moderate level of agreement.

%%%%%%%%%%%%%%%%%%%%%%%%%%%
% \shc{We commented out a paragraph here. Convince us to put it back.}
%%%%%%%%%%%%%%%%%%%%%%%%%%%
% Proper interpretation of this value is somewhat subjective.
% For comparison, we note the example given by~\citet{kripps-alpha}
% in which English-speaking annotators where tasked with assigning television
% characters to categories with Dutch names.
% On this task, annotators also obtained $\alpha=.44$,
% which is coincidentally close to the value our annotators achieved
% on the topic-word matching task.

We note that when using crowdsourcing,
particularly with subjective tasks such as topic-word matching,
we expect somewhat low inter-annotator agreement.
However, previous work indicates that when properly aggregated,
we can still filter out noisy judgments and obtain reasonable opinions
\cite{crowdsource-reliablity}.

%Is the top whisker for LDA the right length? I thought if there were suspected outliers, the whisker was supposed to be 1.5 times the length of IQR.
%I think it might be just that for something to be an outlier it needs to be that far outside of the 
%IQR or the top whisker or something.

\begin{figure}[h]
    \centering
    \includegraphics[width=.50\textwidth]{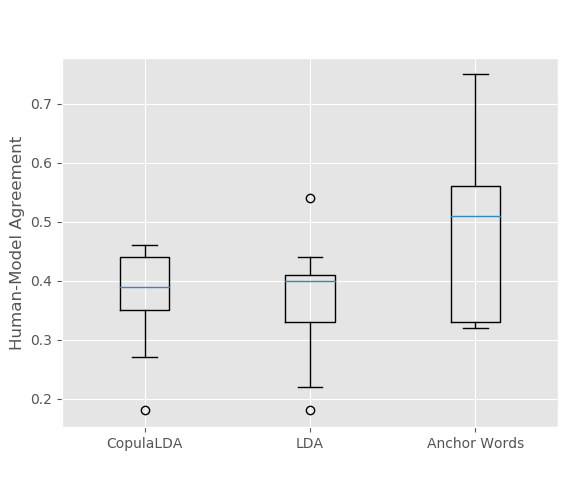}
    \caption{Plot showing human agreement with each model type. CopulaLDA
    performs slightly worse than LDA. Humans preferred topic assignments from
    Anchor Words by a wide margin.}%\jbg{Y-Axis font a little small. Make sure this is PDF}}
    \label{me:fig:model-human}
\end{figure}

Figure~\ref{me:fig:model-human} summarizes the human agreement with the three
different model types.
Surprisingly, despite claiming to produce superior local topic quality,
% and despite performing better than LDA in terms of perplexity
% (a global measure of topic quality),
CopulaLDA actually performs slightly worse than LDA
according to our results with the topic-word matching task.
The fact that CopulaLDA performs poorly despite being designed to improve local topic quality
illustrates the need for effective local topic quality metrics.

We also note that users agree with Anchor Words more often
than LDA by a wide margin, indicating that Anchor Words achieves superior 
token-level topic assignments.
However,
in terms of global topic quality, Anchor Words is roughly similar to
LDA~\cite{anchors-practical}.
One possible explanation for this is that when using Anchor Words the task of 
learning the global topic-word distributions is separate from the problem of 
producing accurate local topic assignments, making both tasks easier.
For many tasks an argument can be made for a joint-model,
so further investigation into this phenomenon is warranted.

\subsection{Global Metrics Correlation}
For Coherence and Significance, we compute a least-squares regression for
 human-model agreement on the topic-word matching task.
As seen in Table~\ref{me:tab:rsquaredglobal},
we report the coefficient of determination ($r^2$) for each global metric and dataset.
Note that global metrics do correlate somewhat with human judgment of local topic quality.
However, the correlation is moderate to poor, especially in the case of coherence,
and we propose new metrics that will achieve greater correlation with human evaluations.

\begin{table}[h]
    \scriptsize    
    \centering
    \begin{tabular}{l l c c c}
    \hline\hline
    & Metric & Amazon & Newsgroups & New York Times\\
    \hline
    \multirow{3}{*}{Global}
    & \abr{SigVac} & 0.6960 & 0.6081 & 0.6063 \\
    & \abr{SigUni} & 0.6310 & 0.4839 & 0.4935 \\    
    & \abr{Coherence} & 0.4907 & 0.4463 & 0.3799 \\
    % & \abr{SigDoc} & 0.0655 & 0.1121 & 0.0798 \\
    \end{tabular}
    \caption{Coefficient of determination ($r^2$) between global metrics and
    crowdsourced topic-word matching annotations.}
    \label{me:tab:rsquaredglobal}
\end{table}

\section{Proposed Metrics}
\label{me:sec:propose}
%If we wanted to add something more visually compelling, an image clarifying the intuition behind the metrics would be a good place for this.

%As previously mentioned,
%recent models, such as CopulaLDA~\cite{cop-lda},
%which claim to improve the quality of token-level topic assignments,
%have only been evaluated using global topic metrics.
We develop an automated methodology for evaluating local topic model
quality.
Following the pattern used by~\citet{coherence-automatic} to develop coherence,
we propose a variety of potential metrics that reflect greater
token-level topic quality such as that in Figure~\ref{me:fig:natural-tag}.
As with coherence, we correlate these automated metrics with human
evaluations in order to determine which automated metric yields the most
accurate estimate of local topic quality as judged by human annotators. \\

\noindent\textbf{Topic Switch Percent (\textsc{SwitchP})} 
It is a platitude of writing that a sentence expresses one idea,
and by this logic we would expect the topic assignments in a sentence
or local token cluster to be fairly consistent.
Using this intuition, we propose our first metric
which measures the percentage of times a topic switch occurs relative
to the number of times a switch could occur.
The intuition behind this is that tokens near each other should switch
infrequently, and thus be consistent in expressing a single idea.
In a corpus with $n$ tokens, with $z_i$ being the topic
assignment of the $i$th token in the corpus, and $\delta(i,j)$ being the
Kronecker delta function, we measure this consistency with
\begin{equation}
\frac{1}{n-1} \sum_{i=1}^{n-1} \delta(z_i,z_{i+1}).
\end{equation} \\

\noindent\textbf{Topic Switch Variation of Information (\textsc{SwitchVI})} 
%\jbg{Put intuition first. What does SwitchP fail to capture?}
Following from the intuition from \textsc{SwitchP}, there are times when
a sentence or local cluster could express multiple ideas, which would
result in frequent natural topic switching.
An example of this is in figure~\ref{me:fig:natural-tag} which has a noun phrase
at the beginning referencing P.Diddy, but then switches to talking about
music videos.
Therefore this proposed metric still penalizes topic switches like \textsc{SwitchP},
but penalizes less those models which switch consistently between the same (presumably related) topics.
This metric uses variation of information (or VI), which measures the amount of information
lost in changing from one partition to another~\cite{vi}.

Assuming that our model has $K$ topics,
and once again using $z_i$ as the topic assignment for token $w_i$,
we consider two partitions
$S=\{S_1,...,S_K\}$ and 
$T=\{T_1,...,T_K\}$
of the set of tokens $w$,
such that 
$S_i = \{w_j | z_j = i\}$ and
$T_i = \{w_j | z_{j+1} = i\}$.
Variation of information is defined as
\begin{equation}
    H(S) + H(T) - 2 I(S, T),
\end{equation}
where $H(\cdot)$ is entropy and $I(S, T)$ is the mutual information between $S$ and $T$.
In other words, we measure how much information we lose in our topic assignments
if we reassign every token to the topic of the token that follows.\\

\noindent\textbf{Average Rank (\textsc{AvgRank})}
The most common way of presenting topics to humans is as a set of related words,
namely the most probable words in the topic-word distributions.
Consequently, we would expect words in the same topic to also occur close
to one another with high frequency.
Leveraging this intuition, where $\rank(w_i, z_i)$ is the rank of $i$th word $w_i$
in its assigned topic $z_i$ when sorted by probability, we use the following:
\begin{equation}
    \frac{1}{n} \sum_{i=1}^n \rank(w_i, z_i).
\end{equation}

With this evaluation the lower bound is 1, although this would require 
that every token be assigned to a topic for which its word is the mode.
However, this is only possible if
the number of topics is equal to the vocabulary size. \\

\noindent\textbf{Window Probabilities (\textsc{Window})} 
Modifying slightly the intuition behind \textsc{SwitchP}
pertaining to local tokens having similar topic assignments,
\textsc{Window} seeks to reward topic
models which have topic assignments which not only explain individual
tokens, but also the tokens within a window around the assignment.
Given a window size, 
and once again using $\phi$ as the topic-word distributions,
we compute the following:
\begin{equation}
    \frac{1}{n(2s+1)}\sum_i^n \sum_{j=i-s}^{i+s} \phi_{z_j, w_i} \;.
\end{equation}
In our experiments, we use a window size of 3 ($s=1$), meaning
that for each token we consider its topic assignment,
as well as the topic assignments for the tokens
immediately preceding and following the target token.
We choose $s=1$ because we want to maintain consistency
while allowing for topics to switch mid-sentence in a natural way.

\noindent\textbf{Topic-Word Divergence (\textsc{WordDiv})} 
% \textsc{SwitchP} and \textsc{SwitchVI} do not take into account the actual probabilities
% in the topic-word distributions.
Stepping away from human intuition about the structure of sentences and topics,
we imagine a statistical approach that explores how the assignments
in a document and the actual word-topic distributions are related.
Given this, consider a topic model with $K$ topics,
$V$ token types, and $D$ documents with topic-word distributions given by a $K \times V$
matrix $\phi$ such that $\phi_{i,j}$ is the conditional probability
of word $j$ given topic $i$.
Furthermore, let $\theta_d$ be the $K$-dimension 
document-topic distribution for the $d$th document
and $\psi_d$ be the $V$-dimensional distribution of words for document $d$.
This metric measures how well the topic-word probabilities explain
the tokens which are assigned to those topics:
\begin{equation}
    \frac{1}{D}\sum_d^D \jsd(\theta_d \cdot \phi \: || \: \psi_d)
\end{equation}
where $\jsd(P\:||\: Q)$ is the Jensen-Shannon divergence between
the distributions $P$ and $Q$.
This evaluation rewards individual topic assignments
which use topics that explain the cooccurrences of an entire
document rather than individual tokens. \\

\section{Automated Evaluations}
\label{me:sec:results}

%We now turn our attention to the correlation between the human judgments
%obtained in Section~\ref{me:ssec:evals} and the automated evaluations
%proposed in Section~\ref{me:sec:propose}.
As before, for each of our proposed metrics, we compute a least-squares regression for
both the proposed metric and the human-model agreement on the topic-word
matching task.
As seen in Table~\ref{me:tab:rsquared},
we report the coefficient of determination ($r^2$) for each metric and dataset.

\begin{table}[h]
    \scriptsize    
    \centering
    \begin{tabular}{l l c c c}
    \hline\hline
    & Metric & Amazon & Newsgroups & New York Times\\
    \hline
    \multirow{5}{*}{Local}
    & \abr{SwitchP}  & 0.9077 & 0.8737 & 0.7022 \\
    & \abr{SwitchVI} & 0.8485 & 0.8181 & 0.6977 \\
    & \abr{AvgRank} & 0.5103 & 0.5089 & 0.4473 \\
    & \abr{Window} & 0.4884 & 0.3024 & 0.1127 \\
    & \abr{WordDiv} & 0.3112 & 0.2197 & 0.0836 \\
    \hline
    \multirow{3}{*}{Global}
    & \abr{SigVac} & 0.6960 & 0.6081 & 0.6063 \\
    & \abr{SigUni} & 0.6310 & 0.4839 & 0.4935 \\    
    & \abr{Coherence} & 0.4907 & 0.4463 & 0.3799 \\
    % & \abr{SigDoc} & 0.0655 & 0.1121 & 0.0798 \\
    \end{tabular}
    \caption{Coefficient of determination ($r^2$) between automated metrics and
    crowdsourced topic-word matching annotations. We include metrics measuring
    both local topic quality
    and global topic quality.}
    \label{me:tab:rsquared}
\end{table}

Humans agree more often with models
trained on Amazon reviews than on New York Times.
This likely reflects the underlying data,
since Amazon product reviews tend to be highly focused on specific products
and product features, and the generated topics naturally reflect these products.
In contrast, New York Times data deals with a much wider array of subjects
and treats them with nuance and detail not typically found in product reviews.
This makes the judgment of topic assignment more difficult and subjective.

Notwithstanding the differences across datasets,
\abr{SwitchP} most closely approximates human judgments of local topic quality,
with an $r^2$ which indicates a strong correlation.
This suggests that when humans examine token-level topic assignments,
they are unlikely to expect topic switches from one token to the next,
which fits with what we observe in Figure~\ref{me:fig:natural-tag}.
As evidenced by the lower $r^2$ for \abr{SwitchVI},
even switching between related topics does not seem to line up with human
judgments of local topic quality.

%Can we be more specific about what this correlation is?
Again, there is a correlation between coherence and the topic-word
matching task, although the correlation is only moderate.
Similarly, word-based significance metrics have a moderate correlation with
topic-word matching.
We maintain that these global topic metrics are important measures for topic
model quality, but they fail to capture local topic quality as
\abr{SwitchP} does.

\section{Discussion}
\label{me:sec:discuss}

Considering the intuition gained from the motivating example in
Figure~\ref{me:fig:split-np}, it is not surprising that humans would
prefer topic models which are locally consistent.
% Historically speaking, the fact that many topic models switch between
% topics so frequently is confusing to users. \shc{This should have a reference?}
Thus, our result that \abr{SwitchP} is correlated with human judgments
of local topic quality best parallels that intuition.

We note that our annotators are only shown the topic assignment for a single token
and do not know what topics have been assigned to the surrounding tokens.
Despite this,
our annotators apparently prefer models which are consistent.
While the result is intuitive,
it is surprising that it is illuminated through a task that asks them to only identify the topic
for a single token.

Given our results, we recommend that topic switch percent be adopted as an
automated metric to measure the quality of token-level topic assignments.
We would refer to this metric colloquially as `consistency'
in the same way that PMI scores on the top $n$ words of a topic
are referred to as coherence.
We advocate that future work on new topic models include validation
with respect to topic consistency, just as recent work has included
evaluation of topic coherence.

However, we are careful to point out that topic consistency should not be
used to the exclusion
of other measures of topic model quality.
After all, topic consistency is trivially maximized by minimizing topic
switches without regard to the appropriateness of the topic assignment.
Instead, we advocate that future models be evaluated with respect to global topic
quality (e.g., coherence, significance, perplexity)
as well as local topic quality (i.e., consistency).
These measures, in addition to evaluation of applicable downstream tasks
(e.g., classification accuracy),
will give practitioners the information necessary to make
informed decisions about model selection.

\section{Conclusion}

We develop a novel crowdsourcing task,
which we call topic-word matching,
to illicit human judgments of local topic model quality.
% Contrary to expectation,
% we find that CopulaLDA actually performs worse than
% other models with respect to this task.
We apply this human evaluation to a wide variety of models,
and find that topic switch percent (or \abr{SwitchP})
correlates well with this human evaluation.
We propose that this new metric,
which we colloquially refer to as consistency,
be adopted alongside evaluations of global topic quality
for future work with topic model comparison.

%\section*{Acknowledgements} This work was supported by the NSF grant XXX-XXXXXXX